\documentclass[final,5p,times,twocolumn]{elsarticle}

\usepackage{epsfig}
\usepackage{graphicx}
\usepackage{amsmath}
\usepackage{amssymb}
\usepackage{tabularx}
\usepackage{color}
\usepackage{booktabs}
\usepackage{colortbl}
\usepackage{multirow}
\usepackage{caption}
\usepackage[table]{xcolor}
\usepackage{rotating} 
\usepackage{subcaption} 
\usepackage{algorithm}  
\usepackage{algpseudocode}  
\usepackage{lineno,hyperref}
\modulolinenumbers[5]

\journal{Pattern Recognition}

\begin{document}

\begin{frontmatter}

\title{Learning Efficient, Explainable and Discriminative Representations for Pulmonary Nodules Classification}


\author[mymainaddress]{Hanliang Jiang}

\author[mysecondaryaddress]{Fuhao Shen}
\author[mysecondaryaddress]{Fei Gao\corref{mycorrespondingauthor}}
\cortext[mycorrespondingauthor]{Corresponding author}
\ead{gaofei@hdu.edu.cn}


\author[mythirdaddress]{Weidong Han}

\address[mymainaddress]{Pulmonary and Critical Care Medicine, Sir Run Run Shaw Hospital, 
School of Medicine, Zhejiang University, Hangzhou 310020, China.}
\address[mysecondaryaddress]{School of Computer Science and Technology, Hangzhou Dianzi University, Hangzhou 310018, China.}
\address[mythirdaddress]{Department of Medical Oncology, Sir Run Run Shaw Hospital, College of Medicine, Zhejiang University, Hangzhou, Zhejiang 310016, China.}

\begin{abstract}
Automatic pulmonary nodules classification is significant for early diagnosis of lung cancers. Recently, deep learning techniques have enabled remarkable progress in this field. However, these deep models are typically of high computational complexity and work in a black-box manner. 
To combat these challenges, in this work, we aim to build an efficient and (partially) explainable classification model. Specially, we use \emph{neural architecture search} (NAS) to automatically search 3D network architectures with excellent accuracy/speed trade-off. Besides, we use the convolutional block attention module (CBAM) in the networks, which helps us understand the reasoning process. 
During training, we use A-Softmax loss to learn angularly discriminative representations. 
In the inference stage, we employ an ensemble of diverse neural networks to improve the prediction accuracy and robustness. 
We conduct extensive experiments on the LIDC-IDRI database. Compared with previous state-of-the-art, our model shows highly comparable performance by using less than 1/40 parameters. Besides, empirical study shows that the reasoning process of learned networks is in conformity with physicians' diagnosis. Related code and results have been released at: \url{https://github.com/fei-hdu/NAS-Lung}.
\end{abstract}

\begin{keyword}
pulmonary nodule classification \sep convolutional neural network \sep neural architecture search \sep computer-aided diagnoses \sep convolutional block attention module
\MSC[2010] 00-01\sep  99-00
\end{keyword}

\end{frontmatter}


\section{Introduction}
\label{sec:intro}

Deep learning \citep{krizhevsky2012imagenet} has witnessed striking advances in the area of Computer-Aided Diagnoses (CAD), including medical image analysis \cite{Taj2016CNN}, medical image segmentation \cite{ada2019MRIseg}, and automatic medical reports generation \cite{Li2018Hybrid}, etc. Such deep-learning based CAD systems could help physicians by offering second opinions and flagging concerning areas in the healthcare data  \citep{DLhealth2019nature}. 


The diagnosis of lung cancer also benefits from recent advances in deep-learning based CAD systems. 
Lung cancer has caused a growing number of deaths at present in the world \citep{Siegel2018cancer}. 
However, due to the sheer volume of computed tomography (CT) images, it is time-consuming for doctors to diagnose. Numerous efforts have been made to develop automated pulmonary nodules detection \citep{xie2019PR} and classification \citep{zhu2018deeplung, JIANG2019AEDPN} algorithms. Such algorithms greatly benefit early-stage lung cancer diagnosis. 

In this work, we focus on the pulmonary nodules classification task, i.e. judging whether a candidate nodule is benign or malignant. Traditionally, researchers explore hand-crafted features and use a classifier to predict the category of a nodule \citep{Fan2014Lung}. Recently, researchers are inspired to employ Convolutional Neural Networks (CNNs) for pulmonary nodule classification \citep{anthimopoulos2016lung}. Initially, researchers use 2D CNNs to classify every CT image individually \citep{shen2017multi}. 
Recently, researchers use 3D CNNs and more complex neural architectures to improve the performance \citep{zhu2018deeplung}.

At present, the remarkable progress enabled by deep learning is mostly due to the complicated network architectures, which are manually developed by human experts. The design of neural architectures is typically time-consuming and relies heavily on experts' experiences. Besides, existing work consider little about the efficiency of networks. In practical applications, there is often a huge volume of medical data but a limited computation budget in most hospitals. It is meaningful to achieve a balance between precision and efficiency. 

In addition, existing networks typically works in a black-box manner. It is significant to develop explainable models for clinical applications \cite{tjoa2020XAImedical}. 
Recently, several efforts have been made for exploring explainable CAD algorithms by using relational learning techniques \cite{jing2020relational}. For example, Xu et al. \cite{xu2019end} used the medical knowledge graph to develop relational dialogue system for automatic diagnosis. Yang et al. \cite{yang2020relational} used relational learning between multiple pulmonary nodules for boosting the prediction accuracy. Zhang et al. \cite{zhang2017mdnet} established a multimodal mapping between medical images and diagnostic reports, and used symptom descriptions and visualize attention to justify the network diagnosis process of pathology bladder cancer. However, either knowledge graph generation or diagnostic collection is of high cost. It is still challenging to directly learn the relationships between the visual patterns in CT images and the expert knowledge in physicians' judgement, for the nodule classification task. 

To combat these challenges, in this paper, we first use automatic \emph{Neural Architecture Search} (NAS) technique \citep{NAS2019JMLR} to design 3D network architectures with excellent accuracy/speed trade-off. 
 Automatic NAS is to find optimal neural architecture from a search space by using some search strategy and performance estimation strategy. Already by now, NAS methods have outperformed manually designed architectures on various tasks, including medical image diagnoses \cite{FAES2019AutoDLmedical} and 3D medical image segmentation \cite{kim2019scalable, weng2019nasunet}. To achieve a best balance between precision and efficiency, we use an advanced NAS method, termed \emph{Partial Order Pruning} (POP) \citep{li2019partial}, in this paper. 

To justify the network diagnosis process, we additionally use the convolutional based attention module (CBAM) \citep{Woo2018CBAM} in the proposed networks. The attention modules in CBAM will characterize the relationships between visual patterns and symptom descriptions.
To improve the margin between the learned representations of malignant and benign lesions, we use A-Softmax loss \citep{Liu2017SphereFace} to train the networks. The A-Softmax loss has shown inspiring performance in enabling CNNs to learn angularly discriminative features.
In the inference stage, we employ an ensemble of divergence neural architectures to further improve the prediction robustness.
Extensive experiments are conducted on the LIDC-IDRI database. The corresponding results show that our final model is highly comparable with previous state-of-the-art (SOTA), but with less than 1/40 parameters. Besides, empirical study shows that the reasoning process is in conformity with physicians' diagnosis. 

Our contributions are mainly four-fold:
\begin{itemize}
\item First, to our best knowledge, this is the first attempt that uses NAS for pulmonary nodules classification;

\item Second, we analyse the reasoning process of the network, which is in conformity with physicians' diagnosis;

\item Third, we employ A-Softmax loss to train the network for learning discriminative representations;

\item Forth, our model is highly comparable with previous SOTA method but with using less than 1/40  parameters. The related code and models have been released at: \url{https://github.com/fei-hdu/NAS-Lung}.
\end{itemize}

The rest of this paper is organized as follows. Section \ref{sec:related} introduces related works. Section \ref{sec:method} details the proposed method. Experimental results and analysis are presented in section \ref{sec:exp}. Section \ref{sec:conclusion} concludes this paper.

\section{Related Work}
\label{sec:related}

In this section, we briefly introduce related works about pulmonary nodule classification and NAS. Please refer to \citep{NAS2019JMLR} for a comprehensive survey of NAS methods.

\subsection{Pulmonary nodule classification}
\label{ssec:related-pnc}

Due to the increasing threat of lung cancer, reaseachers have contribute greatly to develop automated pulmonary nodule classification algorithms \cite{fakoor2013using}. Like other computer vision fields, researchers traditionally use hand-crafted features, e.g. Gabor, Local Binary Patterns (LBP), and SIFT descriptor etc. \cite{Fan2014Lung} to represent a nodule, and then use a \textit{shallow} machine learning techniques, e.g. Support Vector Machine (SVM) \cite{Netto2012LungSVM} and Random Forest \cite{Lee2010Random}, to inference the type of a nodule.

Recently, deep CNNs have achieved great success in various computer vision tasks, such as image classification, segmetnation, and enhancement. Researchers are therefore inspired to classify nodules by using CNNs. Initially, researchers treat each CT frame seperately and use 2D CNNs to learning a classifier. For example, Shen et al. proposed to use a multi-crop CNN \cite{shen2017multi} to make the model robust to scales of nodules. 

Since nodules are naturally 3D, Yan et al. \cite{yan2016classification} explored 3D CNNs for pulmonary nodule classification. Latterly, Zhu et al. \cite{zhu2018deeplung} used 3D deep dual path networks (DPNs) and employ a number of auxiliary features to boost the diagnosis performance. Jiang et al. \cite{JIANG2019AEDPN} sequentially deployed a contextual attention module and a spatial attention module to 3D DPN to improve the representation ability. Besides, they employ an ensemble of different model variants to improve the prediction robustness. These works show inspiring results. 

Although the aforementioned deep models show inspiring results, they are typically of high computational complexity and work in a black-box manner. In this paper, we propose to explore efficient, explainable and discriminative networks by using NAS, CBAM, and A-Softmax loss. As will shown in the experimental section, our resulting model performs competitively with previous SOTA by using less than 1/40 parameters.

\subsection{Neural Architecture Search}
\label{ssec:related-nas}

The success of deep learning in various tasks is accompanied by increasingly more complex neural architectures. Such architectures are manually designed by experts. The designing process is typically time-consuming, and the capacity of networks are bounded by experts' experiences. Consequently, there is a rising demand of automatic Neural Architecture Search (NAS) \citep{NAS2019JMLR}. Already by now, great efforts have been made in three dimensions: search space, search strategy, and performance estimation strategy. Besides, NAS methods have outperformed manually designed architectures on various tasks \citep{real2019aging}. 

Inspired by successes of NAS, several attemps have been made to develop automatic CAD systems by using NAS methods. To name a few, Faes et al. \cite{FAES2019AutoDLmedical} used Google Cloud AutoML to develop medical image diagnostic classifiers for five common diseases. Most NAS models show comparable performance and diagnostic properties to previous SOTA algorithms. Besides, several NAS based 3D medical image segmentation methods are proposed in \cite{kim2019scalable, weng2019nasunet}. Recently, Yu et al. \cite{yu2020c2fnas} proposed a coarse-to-fine neural architecture search (C2FNAS) for 3D medical image segmentation. In C2FNAS, they first search the macro-level topology of the network and then search at micro-level for operations in each cell. Zhang et al. \cite{zhang2019identify} explored hybrid spatio-temporal neural architecture search for fMRI data related tasks.

\begin{figure*}
\centering
\includegraphics[width=0.8\linewidth]{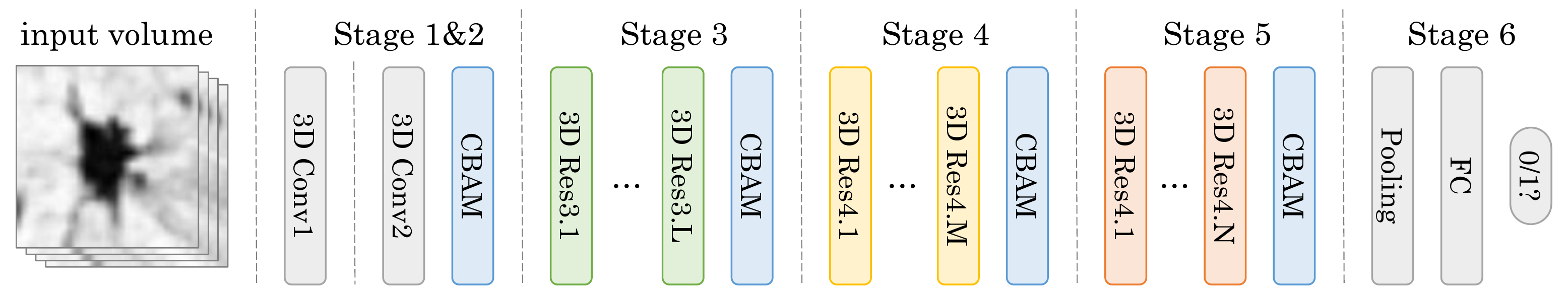} \\
\caption{General 3D network architecture. Here, $L$, $M$ and $N$ are learned in the NAS algorithm.}
\label{fig:netarc}
\end{figure*}

All the previous NAS based medical image analysis method shows inspiring and promising performance. However, the searching procedure typically cost a long time, and the searched networks are large-size. In this paper, we aim to search 3D neural networks with the best accuracy/speed trade-off. 

\section{Proposed}
\label{sec:method}
In this paper, we use the 3D NAS method, CBAM module, A-Softmax loss, and ensemble strategy to learn efficient, explainable and discriminative representations for pulmonary nodules classification.  
In this section, we will sequentially detail these techniques in the following.

\subsection{General 3D Network Architecture}
\label{ssec:netarc}
Our general 3D network architecture is as shown in Fig. \ref{fig:netarc}. 
Stages 1 and 2 include one 3D convolutional layer, i.e. Conv1 and Conv2, respectively. The outputs  of both stage 1 and stage 2 have 4 channels. 
Stages 3-5 include $L, M, N$ residual blocks, respectively. The number of channels of the $i$-th residual block in stage $s (s=3,4,5)$ is denoted by $C^s_i$. We will learn optimal $L, M, N$, and $C^s_i$ in the searching process by using NAS (Section \ref{ssec:3dnas}). 
Stage stage 6 produces the final prediction with a global average pooling and a fully-connected (FC) layer. The FC layer outputs a binary label denoting whether a input nodule is benign or malignant.
We expand a CBAM \citep{Woo2018CBAM} to the last layer of Stages 2-5, respectively. 

\begin{figure}
	\centering
	\includegraphics[width=0.9\linewidth]{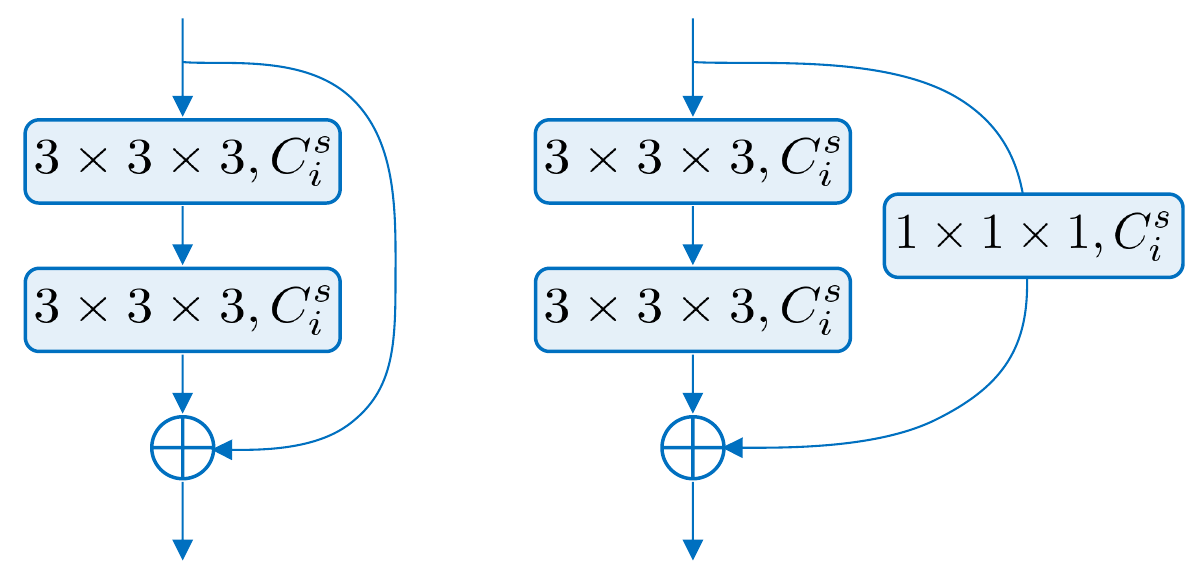} \\
	\caption{The 3D residual blocks used throughout this paper.}
	\label{fig:resblock}
\end{figure}

As shown in Fig. \ref{fig:resblock}, our residual block consists of two 3D convolution layers and a short-cut connection. In case if the size of input does not match the output, an additional 3D convolutional layer is added (as shown in the right sub-figure). Batch normalization (BN) and ReLU nonlinearity are used after all convolutional layers.

\subsection{Convolutional Block Attention Module (CBAM)}
\label{ssec:cbam}
In the general networks, we expand CBAM \citep{Woo2018CBAM} to the basic residual 3D networks, as shown in Fig. \ref{fig:netarc}. CBAM includes a channel attention module and a spatial attention module. Such attention modules can help us pry into the networks and partially understand the reasoning process. Besides, as will shown in Section \ref{ssec:exp_cbam}, CBAM boosts the classification accuracy. 

\begin{figure*}
\centering
\includegraphics[width=0.6\linewidth]{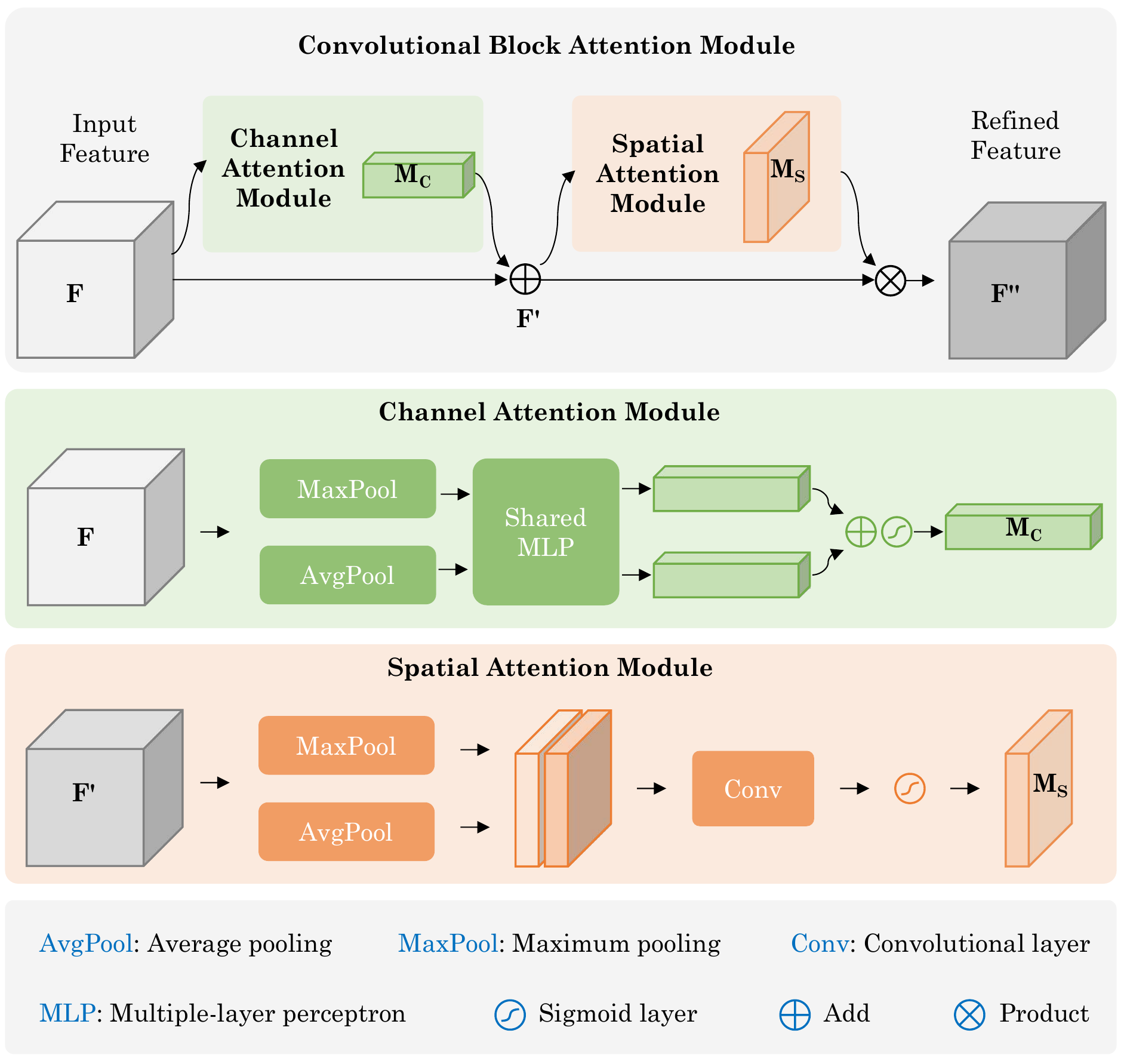} \\
\caption{Pipeline of convolutional block attention module (CBAM) \citep{Woo2018CBAM}.}
\label{fig:cbam}
\end{figure*}

As shown in Fig. \ref{fig:cbam}, given a feature map $\mathbf{F}$, the channel attention module infers an 1D channel attention vector $\mathbf{M_C}$, indicating the significance of each channel. The spatial attention model infers a 3D spatial attention map $\mathbf{M_S}$, indicating the significance of each location. The overall process is operated as:
\begin{equation}
\begin{aligned}
\mathbf{F}' = \mathbf{M_C}(\mathbf{F}) \odot \mathbf{F}, \\
\mathbf{F}'' = \mathbf{M_S}(\mathbf{F}') \odot \mathbf{F}',
\end{aligned}
\end{equation}
where $\odot$ denotes element-wise multiplication.

As shown in Fig. \ref{fig:cbam}, the channel attention is computed by:
\begin{equation}
\mathbf{M_C}(\mathbf{F})=\sigma(\mathrm{MLP}(\mathrm{AvgPool}(\mathbf{F})) + \mathrm{MLP}(\mathrm{MaxPool}(\mathbf{F}))),
\end{equation}
and the spatial attention is computed by: 
\begin{equation}
\mathbf{M_S}(\mathbf{F})=\sigma(\mathrm{Conv}([\mathrm{AvgPool}(\mathbf{F}); \mathrm{MaxPool}(\mathbf{F})])).
\end{equation}
Here $\sigma(\cdot)$ denotes the Sigmoid function; $\mathrm{MaxPool}(\cdot)$ and $\mathrm{AvgPool}(\cdot)$ denotes max-pooling and average-pooling, respectively; $\mathrm{MLP}(\cdot)$ denotes a multi-layer perceptron (MLP), which concludes two fully-connected layers followed with a ReLU layer, respectively; $\mathrm{Conv}(\cdot)$ denotes a 3D convolutional layer.

\subsection{A-Softmax Loss Function}
\label{ssec:obj}
Finally, we use the angular softmax (A-Softmax) loss to train the whole network for learning angularly discriminative features \citep{Liu2017SphereFace}. The A-Softmax Loss is expressed as:
\begin{equation}
L_{ang}=\frac{1}{N}\sum_{i}-\log(\frac{e^{||\mathbf{x}_i||\cos(m\theta_{y_i,i})}}{e^{||\mathbf{x}_i||\cos(m\theta_{y_i,i})} + \sum_{j\not ={y_i}}e^{||\mathbf{x}_i||\cos(\theta_{j,i})}})
\label{eq:loss}
\end{equation}
where $\theta_{j,i} (0 \leq \theta_{j,i} \leq \pi)$ is the angle between vector $\mathbf{W}_j$ and $\mathbf{x}_i$; $\mathbf{x}_i$ and $y_i$ are the $i$-th training sample and the corresponding ground-truth label; $\mathbf{W}_j$ and $\mathbf{W}_{y_i}$ are the $j$-th and $y_i$-th column of $\mathbf{W}$, respectively; $\mathbf{W}$ denotes parameters in the last fully-connected layer; $m$ quantitatively controls the size of angular margin. In the implementation, we set $m=4$.

\subsection{Efficient Neural Architecture Search}
\label{ssec:3dnas}

In this paper, we employ the \emph{Partial Order Pruning} (POP) \citep{li2019partial} algorithm to design 3D networks with excellent speed/accuracy trade-off. The basic assumption of POP is that: \textit{a narrower network is always more efficient and less accurate than a wider one}. Following this assumption, Li et al. consider both accuracy and inference speed in the objective function, and use a cutting plane algorithm to solve the corresponding optimization problem. 

In the implementation, we use the general 3D network architecture shown in Table \ref{fig:netarc} in our search space. 
Due to the computation budget limitation, we reduce the search space by restricting:
\begin{equation}
\label{eq:space}
\lfloor(L+M+N)/4\rfloor \leq D \leq \lceil (L+M+N)/2 \rceil, \forall D \in \{L, M, N\}.
\end{equation} 
Unlike the settings in \citep{li2019partial}, we aim at searching for low-latency neural architectures in this paper. In practice, we set the search space as: 
\begin{equation}
C_i^s \in \{4, 8, 16, 32, 64, 128\}, 3\leq L+M+N\leq 9.
\end{equation}
Consequently, all the candidate models would include no more than 9 residual blocks, and each residual block contains no more than 128 channels. With these settings, we significantly narrow down the search space. 
The searched network architecture is denoted by:
\begin{equation}
[[C^3_1, ..., C^3_L], [C^4_1, ..., C^4_M], [C^5_1, ..., C^5_N]].
\label{eq:netarc}
\end{equation}

POP iteratively searches for the networks with the best accuracy/speed trade-off in the whole search space. Each time POP trains a new architecture and obtains its accuracy, and then updates the pruned search space. Here the pruned search space includes the networks which have higher latency but lower accuracy. These architectures are unlikely to provide better speed/accuracy trade-off than the trained networks. By pruning these architectures from the search space, POP avoids unnecessary training cost and thus speeds up the architecture search process. The search process is stopped if the no change to the search space happens for several iterations. For details about the POP algorithm, please refer to the original literature \citep{li2019partial}. 

\textbf{Note:} In default, we has CBAM in our general networks and search for the best depth and width, by training each network using the A-Softmax loss. We may also use the general networks without CBAM in the search space, and first train each network using the Softmax loss. Afterwards, we select a number of networks with excellent  accuracy/speed trade-off, and then expand CBAM blocks to them and train them using the A-Softmax loss. Our experimental result shows that we can obtain models with similar accuracy in both manners. However, using both CBAM and the A-Softmax loss in the search process dramatically reduces the latency.  


\subsection{Ensemble Inference}
\label{ssec:inference}

In the inference stage, we fuse the outputs of different neural architectures to get the final prediction  \citep{DeepForest}. In practice, suppose we select $n$ networks which shows good accuracy/speed trade-off. Each network produces a probability that whether an input nodule is malignant or not. We transfer it to a binary label (i.e. $\{1, 0\}$) by using a threshold of $0.5$. If more than $\lfloor n/2 \rfloor$ models output a label $1$ (i.e. malignant), the input nodule is estimated as malignant (i.e. positive). In the following, we refer to the full model as \texttt{NASLung}.


\section{Experiments}
\label{sec:exp}

We conduct a series of experiments to verify the proposed method and empirically analyse the reasoning process. We will present the experimental settings and analyse the corresponding results below. 

\subsection{Settings}
\label{ssec:exp_settings}

\subsubsection{Dataset} 
\label{ssec:exp_data}
In the experiments, we use the LIDC-IDRI dataset \citep{armato2015LIDCdata, armato2011LIDCpub} with the LUNA16's settings \citep{kuan2017deep}. Specially, the CTs with slice thickness greater than 3mm, slice spacing inconsistent or missing slices, are removed from the LIDC-IDRI dataset. There are totally 1,004 nodules left, in which 450 nodules are positive. The LUNA16 dataset explicitly gives a 10-fold cross validation split. Correspondingly, in each run, we use 9 folds (including about 900 samples) for training and the left 1 fold (including about 100 samples) for testing. Following the settings in \citep{zhu2018deeplung}, we evaluate our method on folds 1-5, and report the average performance. 

\subsubsection{Performance Criteria} 
\label{ssec:exp_crit}
To evaluate the pulmonary nodule classification performance, we use four widely used indices, including:
\begin{eqnarray}
\label{Eq:Precision}
	\mathrm{Accuracy} & = & \frac{\mathrm{TP+TN}}{\mathrm{TP+FN+TN+FP}},\\
    \mathrm{Sensitivity} & = & \frac{\mathrm{TP}}{\mathrm{TP+FN}},\\    
    \mathrm{Specificity} & = & \frac{\mathrm{TN}}{\mathrm{FP+TN}}, \text{and}\\
    \mathrm{F1~Score} & = & \frac{\mathrm{2 \cdot TP^2}}{\mathrm{2TP+FP+FN}},
\end{eqnarray}
where, TP, FN, FP, and TN sequentially denote true positive, false negative, false positive, and true negative. Greater values of these criteria indicate better performance. F1 Score evaluates the trade-off between Sensitivity and Specificity. In general, the highest F1 Score indicates the best performance.

\subsubsection{Implementation Details} 
\label{ssec:train}
We implement our framework by using \texttt{Pytorch} and one GTX 1080Ti GPU. In the training procedure, we pad the nodules of size $32 \times 32 \times 32$ into $36 \times 36 \times 36$, and then randomly crop $32 \times 32 \times 32$ from it. We use horizontal flip, vertical flip, z-axis flip for data augmentation. We use Adam optimizer with a learning rate $0.0002$, and momentum parameters $\beta_1=0.5$ and $\beta_2=0.999$. 

\begin{figure}
\centering
\includegraphics[width=1\linewidth]{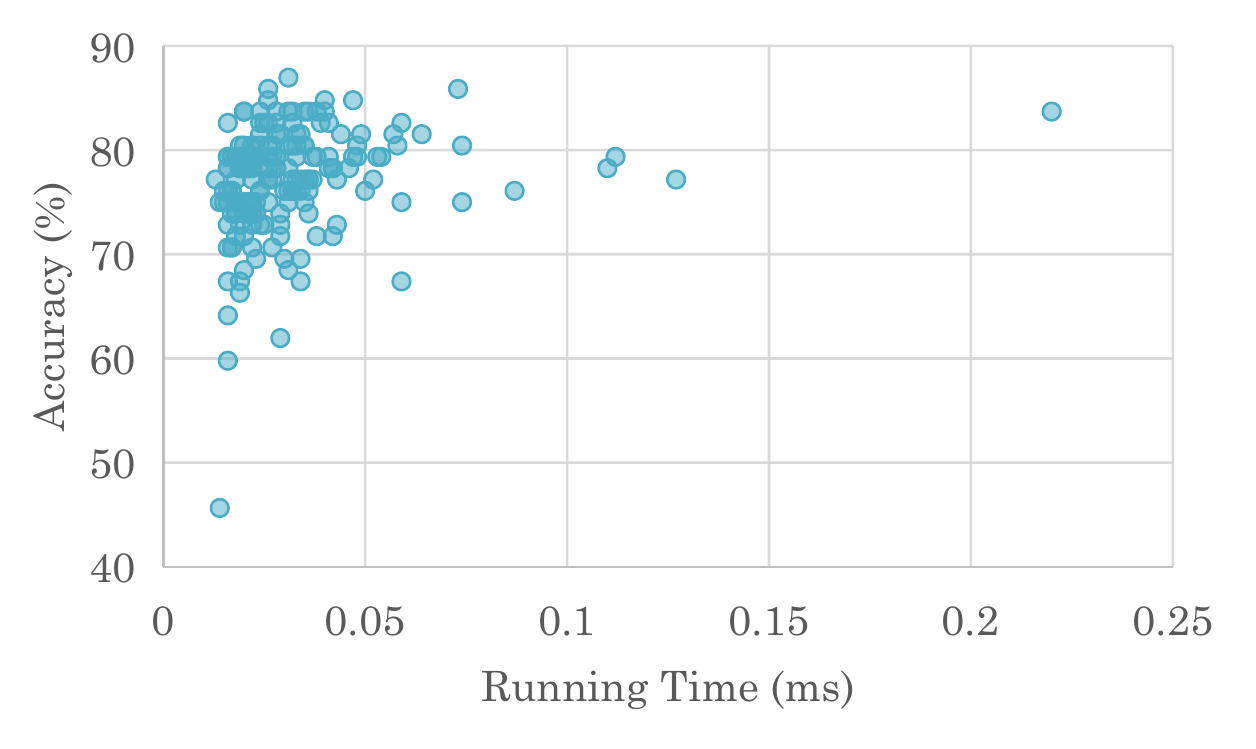} \\
\caption{The accuracy and inference speed of neural models in the search space. Each model is trained for $5$ epochs.}
\label{fig:space}
\end{figure}

In the NAS process, we train each candidate model for $10$ epochs. The search process costs about 8 hours. Fig. \ref{fig:space} shows the accuracy and speed of neural models in the search space. Obviously, most models show inspiring accuracy and high inference speed. Some of them even achieves an accuracy over $86\%$ but cost less than $0.4 ms$ for each nodule. Such excellent candidates allow us to develop real-time automatic classification models. We then select a number of good models according to both \textit{accuracy} and \textit{inference speed}. For each selected model, further optimize it for $700$ epochs by using the A-Softmax loss. Our final ensemble model concludes 9 light-weight networks. 


\subsection{Comparison with Existing Works}
\label{ssec:exp_comp}

We first compare our method with several existing advanced methods, including Multi-crop CNN \citep{shen2017multi}, Nodule-level 2D CNN \citep{yan2016classification}, Vanilla 3D CNN \citep{yan2016classification}, DeepLung \citep{zhu2018deeplung}, and AEDPN \citep{JIANG2019AEDPN}. Here, we report the performance of our ensemble model, NASLung, in Table \ref{tab:pfm_comp}. 

Obviously, our NASLung achieves the highest Accuracy and Specificity, and the second best F1 Score. Both high Accuracy and F1 Score demonstrates that NASLung  trades off well between Sensitivity and Specificity. In other words, NASLung  can correctly classify most nodules, which would dramatically light the burden of physicians. 
Notably, AE-DPN \citep{JIANG2019AEDPN} achieves the best Sensitivity value and F1 Score. However, the model size of AE-DPN is almost 40 times of NASLung. 
All these results demonstrate that by using NAS, CBAM, A-Softmax, and the ensemble strategy, the proposed method achieves an excellent accuracy/efficiency trade-off. 


\begin{table*}
\centering
\caption{Comparison with existing methods across folds 1-5 on the LIDC-IDRI dataset. \textit{Accu.}, \textit{Sens.}, \textit{Spec.}, and \textit{para.} denote \textit{Accuracy}, \textit{Sensitivity}, \textit{Specificity}, and the number of parameters, respectively. The best and second best results in each column are shown in \textbf{boldface} and \underline{underline} format, respectively.}
\small
\label{tab:pfm_comp}
\begin{tabular}{l|ccccc}
\toprule
 			& Accu.	& Sens.	& Spec.	& F1 Score & para. (M)	\\	
\midrule	
Multi-crop CNN \citep{shen2017multi} &  	87.14 	&	--	&	--	&	--	&	--	\\
Nodule-level 2D CNN \citep{yan2016classification} & 	87.30 	&	88.50 	&	86.00	&	87.23 &	--		\\
Vanilla 3D CNN \citep{yan2016classification} & 	87.40 	&	\underline{89.40} 	&	85.20	&	87.25 &	--			\\
DeepLung \citep{zhu2018deeplung} &  	\underline{90.44} 	&	81.42 	&	--	&	--	& \underline{141.57} \\
AE-DPN \citep{JIANG2019AEDPN}  & 	90.24	&	\textbf{92.04}	&	\underline{88.94}	&	\textbf{90.45} & 678.69	\\
\midrule
NASLung (ours) &	\textbf{90.77}	&	85.37 	&	\textbf{95.04} 	&	\underline{89.29} 	&	\textbf{16.84} 	\\
\bottomrule												
\end{tabular}
\end{table*}

%


\subsection{Ablation Study}
\label{ssec:exp_ablation}
We then analyse the impact of the proposed techniques, i.e. CBAM, A-Softmax loss ($\mathcal{L}_{ang}$), and the ensemble strategy. 
We first remove the CBAM blocks from the general architecture, and select the best 5 NAS models as bases. Afterwards, we build a number of model variants by selectively applying CBAM (i.e. NAS+CBAM), A-Softmax loss (i.e. NAS+$\mathcal{L}_{ang}$) or both (i.e. NAS+CBAM+$\mathcal{L}_{ang}$) to them. 
Besides, we conduct experiments by searching the best networks while using CBAM and the A-Softmax loss in NAS. The corresponding results are denoted by NASLung$_{single}$. 
The averages (\textit{avg.}) and standard deviations (\textit{std.}) of F1 Scores of these models under each setting are reported in Table \ref{tab:ablation}.

Obviously, using CBAM or $\mathcal{L}_{ang}$ averagely improves F1 Score by about 0.1 and 0.2, respectively. Using both of them improves F1 Scores by 0.36 in average. In addition, the standard deviation  of F1 Score significantly decreases when we use CBAM, $\mathcal{L}_{ang}$ or both of them. These observations imply that both CBAM and $\mathcal{L}_{ang}$ boost the accuracy and robustness of the original NAS models in general. Besides, the effect of CBAM and $\mathcal{L}_{ang}$ are at least partially additive. 
 Finally, NASLung$_{single}$ shows the best average performance with the lowest standard deviation in general. The average size of NASLung$_{single}$ is about 1/4 of the other models. This comparison implies that using CBAM and the A-Softmax loss in the search space dramatically reduce the latency of networks and slightly improves the accuracy.


\begin{table*}
\centering
\caption{Statistics of performance indices related to different model variants.}
\small
\label{tab:ablation}
\begin{tabular}{l|ccccc}
\toprule
\textit{avg}. & Accu.	& Sens.	& Spec.	& F1 Score & para.(M) \\
\midrule
NAS (base)	&	87.58 	&	84.03 	&	90.50 	&	85.74 	&	7.80 	\\
NAS+CBAM	&	87.61 	&	84.01 	&	90.41 	&	85.86 	&	7.84 	\\
NAS+$\mathcal{L}_{ang}$	&	87.87 	&	82.89 	&	\textbf{91.75} 	&	85.96 	&	7.80 	\\
NAS+CBAM+$\mathcal{L}_{ang}$	&	87.77 	&	\textbf{84.26 }	&	90.60 	&	86.10 	&	7.84 	\\
NASLung$_{single}$	&	\textbf{88.03} 	&	83.39 	&	91.73 	&	\textbf{86.18} 	&	\textbf{1.90}	\\
\bottomrule \toprule
\textit{std.} & Accu.	& Sens.	& Spec.	& F1 Score & para.(M) \\
\midrule
NAS (base)	&	1.06 	&	3.01 	&	\textbf{0.92} 	&	1.50 	&	4.35 	\\
NAS+CBAM	&	0.74 	&	2.15 	&	1.17 	&	1.01 	&	4.37 	\\
NAS+$\mathcal{L}_{ang}$	&	0.62 	&	2.41 	&	2.18 	&	0.73 	&	4.35 	\\
NAS+CBAM+$\mathcal{L}_{ang}$	&	0.56 	&	1.24 	&	1.33 	&	0.56 	&	4.37 	\\
NASLung$_{single}$	&	\textbf{0.44} 	&	\textbf{0.87} 	&	1.04 	&	\textbf{0.49} 	&	\textbf{1.63} 	\\
\bottomrule												
\end{tabular}
\end{table*}

\subsection{Analysis of CBAM}
\label{ssec:exp_cbam}

\begin{figure*}
\centering
\includegraphics[width=0.8\linewidth]{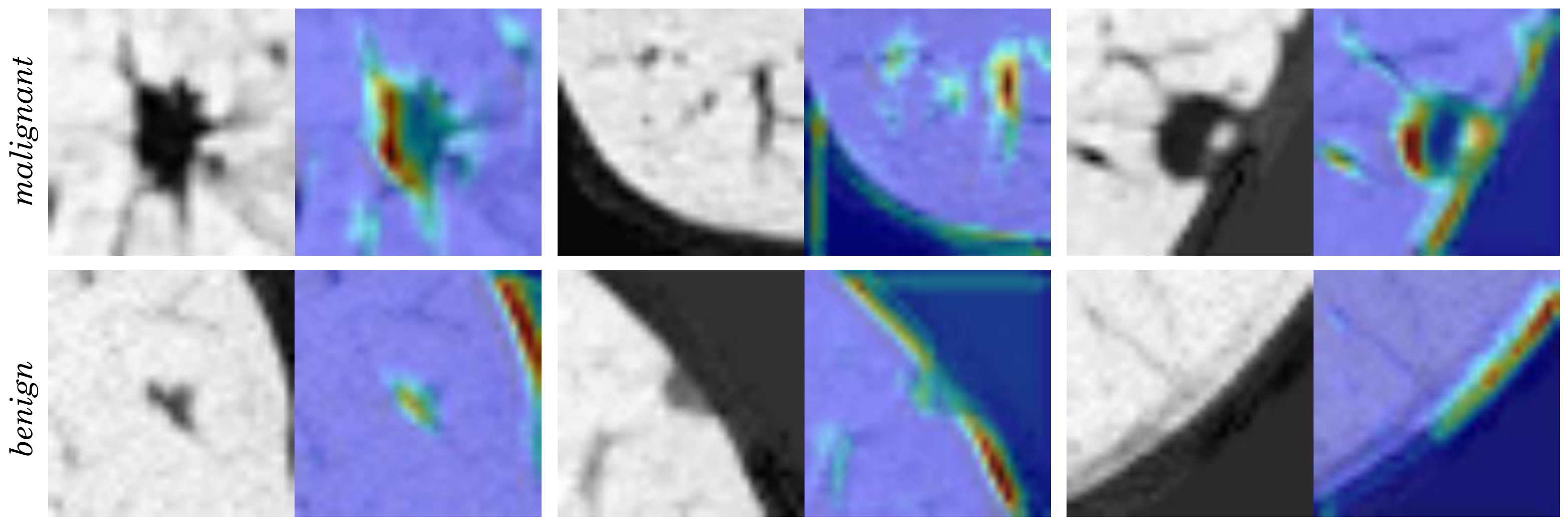} \\
\caption{Visualization of the learned spatial attention. Each example includes two images: the left one is the CT image patch; the right one is the corresponding attention map. Here, red colors indicate high attention values, and blue indicates low attention values.}
\label{fig:visatt}
\end{figure*}

\textbf{Attention Visualization in CBAM.} 
In order to understand the reasoning process, we empirically analyse the relations between the learned spatial attention maps and physicians' diagnosis. To this end, we visualize the learned spatial attention map in Stage 2 due to its small perceptual field. In other words, each attention value here presents the importance of a small region in an input CT image. For the visualization purpose, we only show a 2D slice of the 3D CT cube and the corresponding spatial attention map. 
Fig. \ref{fig:visatt} illustrates several malignant/benign nodules and their corresponding learned attention maps. 

As shown in Fig. \ref{fig:visatt}, greater attention values mainly appear over the borders and inside of nodules. Coincidentally, physicians judge whether a nodule is malignant or benign lesions based on the same areas. Specially, the learned attention maps correspond with physicians' diagnosis in the following three aspects \cite{siegelman1986pulmonary}:
\begin{itemize}
\item First, malignant lesions typically have a lobular border. Statistically, the majority of lesions with a lobular border are malignant. Lobular borders are therefore significant for diagnosis.

\item Second, lobulation leads to heterogeneous growth rates within nodules. Tumour nodules generally grow like lobes (sprouting), not concentrically, so the growing nodules will not be in the center of the lesion. The red area inside a nodule (not located in the center of the lesion) coincides with such eccentric growth patterns. The red area therefore is helpful for the judgement of malignant nodules. 

\item Finally, because lobular borders may also appear with benign lesions, the insides of nodules are more important than borders in clinical diagnosis. Examples illustrated in the top row of Fig. \ref{fig:visatt} show consistent results.
\end{itemize}
These observations indicates that the reasoning process of our model is (at least partially) in conformity with physicians' diagnosis.

\textbf{Arrangement of the channel and spatial attention in CBAM.} 
In addition, we have conducted experiments by changing the sequence of the channel attention with the spatial attention block in Figure \ref{fig:cbam}. We refer to the original and switched versions of CBAM as channel-first order and spatial-first order, respectively. As shown in Table \ref{tab:cbam}, the channel-first order is slightly better than the spatial-first order. This comparison result is exactly the same as that shown in the original literature \cite{Woo2018CBAM}. We thus adopt the channel-first order in our final model. 

\begin{table}
\centering
\caption{Performance about the arrangement of the channel and spatial attention in CBAM..}
\small
\label{tab:cbam}
\begin{tabular}{c|cccc}
\toprule
& Accu.	& Sens.	& Spec.	& F1 Score \\
\midrule	
channel-first & 	\textbf{88.03}	& \textbf{83.39}	& 91.63	& \textbf{86.18} \\
spatial-first & 	87.14	& 80.95	& \textbf{92.15}	& 84.90 \\
\bottomrule												
\end{tabular}
\end{table}

\subsection{Analysis of the A-Softmax loss}
\label{ssec:exp_sphere}

We further analyse the impact of A-Softmax loss in representation learning.
 To this end, we use the Davies-Bouldin Index (DBI) to evaluate the discriminability of the learned features \cite{davies1979cluster}. DBI is a popular metric for evaluating clustering algorithms. Lower values of DBI indicate smaller intra-class distance and larger inter-class distance. Let $X_{i,j}$ denote the feature vector of the $j$-th sample in the $i$-th class. We first compute the intra-class distance by:
\begin{equation}
S_i = \frac{1}{T_i} \sum_{j=1}^{T_i} \Vert X_{i,j} - \bar{X}_i \Vert_2,
\end{equation} 
where $T_i$ is the number of samples of the $i$-th class, $\bar{X}_i$ is the average feature vector (i.e. center) of the $i$-th class; $i=0$ and $1$ indicates the negative and positive classes, respectively. We then compute the iner-calss distance by: 
\begin{equation}
M_{0,1} = \Vert \bar{X}_0 - \bar{X}_1 \Vert_2.
\end{equation} 
Afterwards, DBI is calculated by: 
\begin{equation}
M_{0,1} = \frac{S_0 + S_1}{M_{0,1}}.
\end{equation} 

We calculated these metrics by using the deep features learned with the Softmax loss and A-Softmax loss, respectively. As shown in Table \ref{tab:asoft}, the A-Softmax loss leads to smaller intra-class distance of the negative nodules, larger inter-distance between positive and negative nodules, and smaller DBI, compared to the Softmax loss. Such  dramatic superiority, especially in terms of DBI, implies the learned features by using the A-Softmax loss diverse apparently between positive and negative nodules. In other words, we can learn much more discriminable features by using the A-Softmax loss than using the Softmax loss.

\begin{table}
\centering
\caption{Evaluations of the deep features learned by using the Softmax loss and A-Softmax loss, respectively. Smaller DBI generally indicates smaller intra-class distances (i.e. $S_0$ and $S_1$) and larger inter-class distance (i.e. $M_{0,1}$).}
\small
\label{tab:asoft}
\begin{tabular}{c|cccc}
\toprule
& $S_0$ &	$S_1$	& $M_{0,1}$	& DBI \\
\midrule
A-Softmax loss 	& \textbf{0.439}	& 0.551	& \textbf{0.718}	& \textbf{1.453} \\
Softmax loss	& 0.565	& \textbf{0.515}	& 0.470	& 2.298 \\
\bottomrule												
\end{tabular}
\end{table}

\begin{figure*}
\centering
\includegraphics[width=0.8\linewidth]{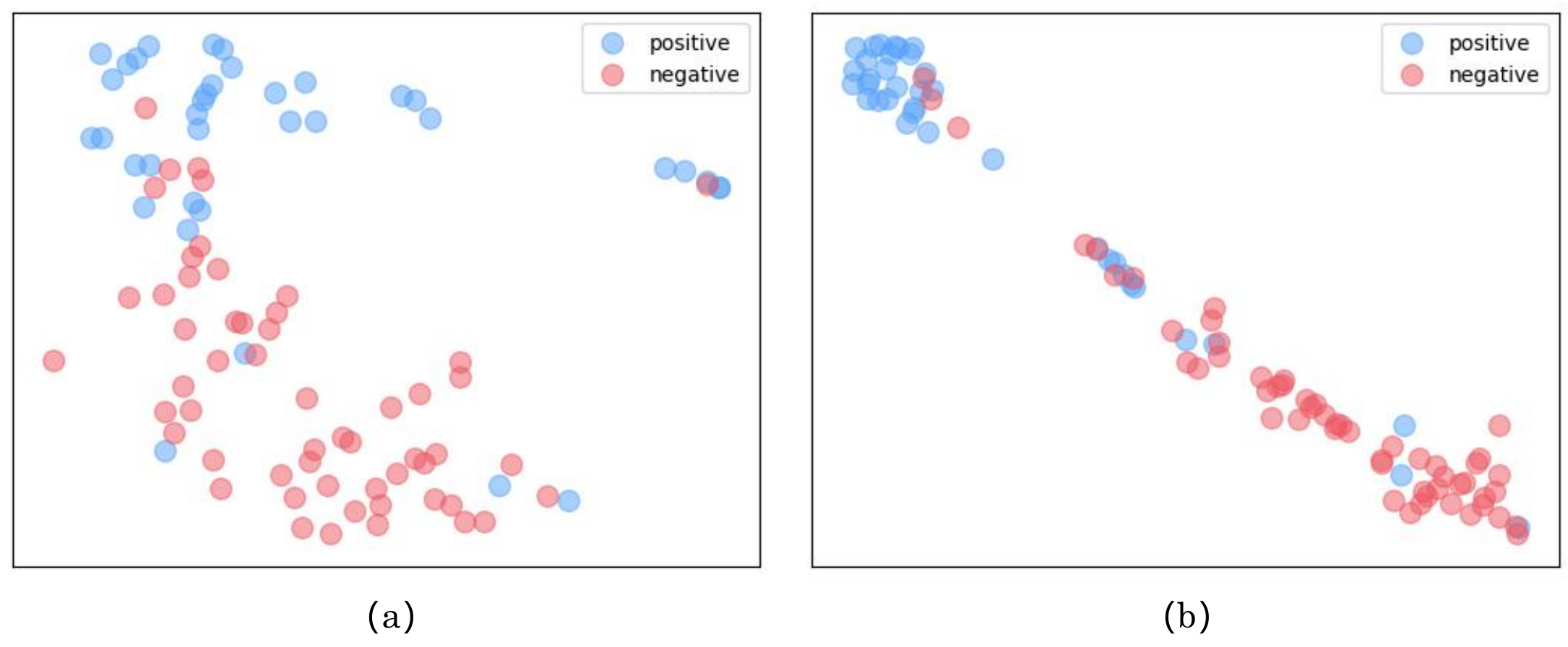} \\
\caption{Visualization of the learned features: (a) the features learned by the Softmax loss and (b) those learned by the A-Softmax Loss. }
\label{fig:tsne}
\end{figure*}

In addition, we visualize the outputs of the penultimate layer in a selected 3D network. Specially, we train hte network by using the Softmax, i.e. Cross-Entropy (CE), loss and A-Softmax loss, respectively. After training, we reduce each feature vector to 2 dimension by using \texttt{t-SNE} \cite{maaten2008tSNE} and visualize them in Fig. \ref{fig:tsne}. In Fig. \ref{fig:tsne}, blue points present positive (malignant) nodules and red points present negative (benign) nodules. We use the \texttt{scikit-learn} toolbox \cite{2011scikitlearn} in the implementation. 

Obviously, compared with the Softmax loss, the A-Softmax loss makes the learned representations more compact intra each class. Besides, the inter-class angular margin in Fig. \ref{fig:tsne}(b) is more distinct than that in Fig. \ref{fig:tsne}(a). Recall that the A-Softmax loss generally improves the classification performance, as previously shown in Table \ref{tab:ablation}. We can safely draw the conclusion that the A-Softmax loss enables networks to learn discriminative representations and benefit the classification performance.

\subsection{Analysis of the Ensemble Strategy}
\label{ssec:exp_ensemble}

\begin{figure*}
\centering
\includegraphics[width=0.8\linewidth]{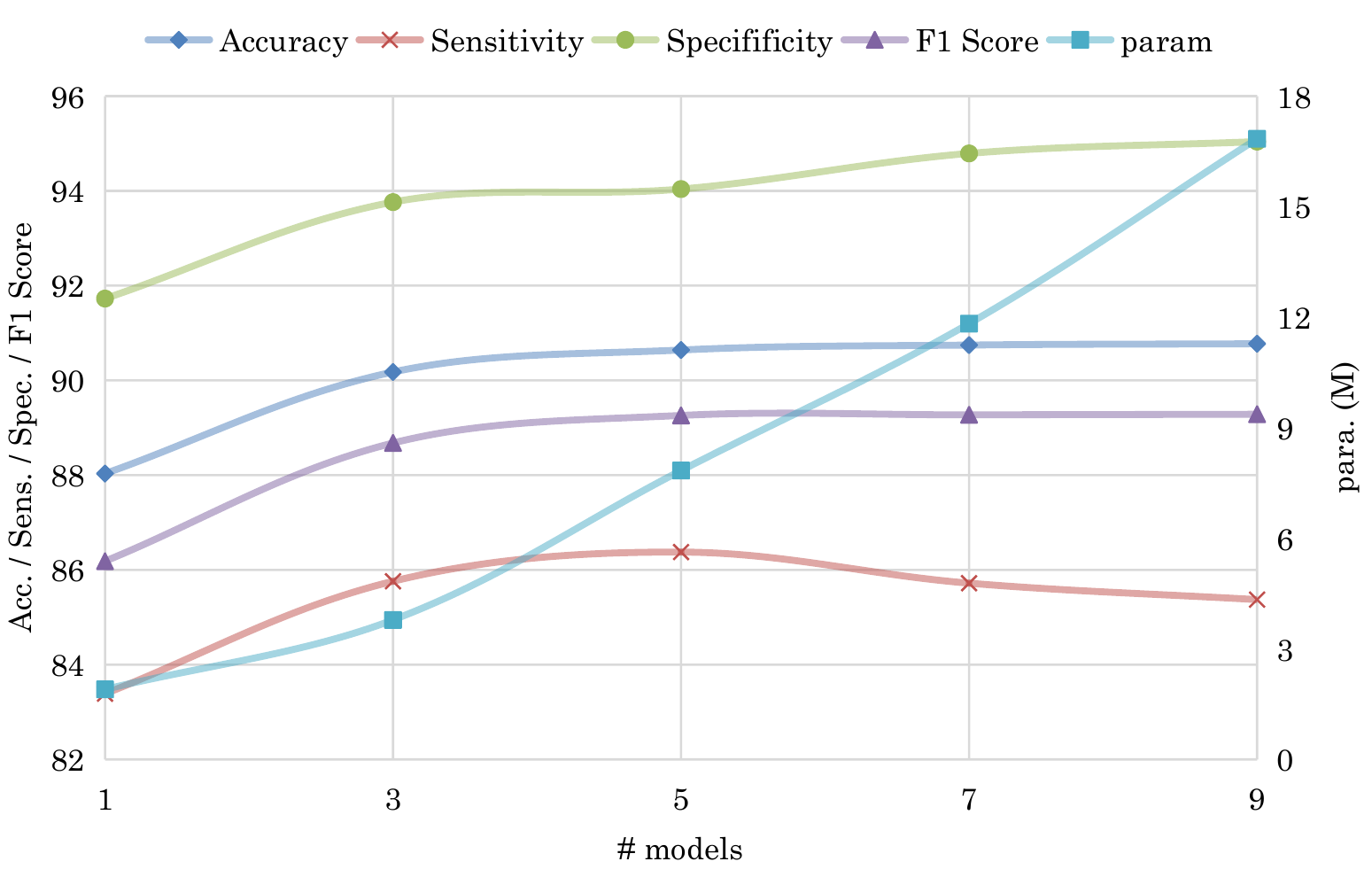} \\
\caption{Performance of the proposed method while using different number of models.}
\label{fig:ensemble}
\end{figure*}

To verify the ensemble strategy, we first randomly select $n$ models for an ensemble prediction, with $n=1, 3, 5, 7, 9$. We repeat this progress for 10 times, and calculate the average performance indices, related to each $n$.  As shown in Fig. \ref{fig:ensemble}, using more models generally improves the performance. Besides, using more than 5 models gains little performance improvement.

%

In addition, we illustrate the performance of the best 10 searched single networks. As shown in Table \ref{tab:ens_cmp}, all these single models achieves a F1-Score higher than 85 and contains no more than 5M parameters each. Remarkably, Model-1 achieves the best F1-Score over 87, by using merely 0.14M parameters. Specially, the architecture of Model-1 is [4,4,[4, 4], [4, 8], [8, 8]]. It is apparent that these single models perform diversely according to different criteria. Integrating all these models would make them complement each other. Accordingly, NASLung outperforms all these single models by a large margin in general. 

\begin{table}
\centering
\caption{Performance of the best 10 searched single networks. \textit{Accu.}, \textit{Sens.}, \textit{Spec.}, and \textit{para.} denote \textit{Accuracy}, \textit{Sensitivity}, \textit{Specificity}, and the number of parameters, respectively. The best and second best results in each column are shown in \textbf{boldface} and \underline{underline} format, respectively.}
\small
\label{tab:ens_cmp}
\begin{tabular}{c|ccccc}
\toprule
 & Accu.	& Sens.	& Spec.	& F1 Score & para.\\	
\midrule	
Model-1	&	\underline{88.83} 	&	\textbf{87.20} 	&	90.12 	&	\underline{87.50} 	&	\textbf{0.14 }	\\
Model-2	&	88.42 	&	84.38 	&	91.46 	&	86.67 	&	2.61 	\\
Model-3	&	88.17 	&	84.44 	&	91.60 	&	86.50 	&	3.90 	\\
Model-4	&	88.13 	&	83.20 	&	92.28 	&	86.30 	&	2.54 	\\
Model-5	&	87.97 	&	83.72 	&	91.31 	&	86.22 	&	0.43 	\\
Model-6	&	87.77 	&	83.67 	&	91.00 	&	86.03 	&	\underline{0.22} 	\\
Model-7	&	87.76 	&	84.14 	&	89.79 	&	85.98 	&	0.86 	\\
Model-8	&	88.00 	&	82.43 	&	92.69 	&	85.97 	&	4.02 	\\
Model-9	&	88.04 	&	78.01 	&	\textbf{96.09} 	&	85.36 	&	4.06 	\\
Model-10	&	87.22 	&	82.70 	&	90.92 	&	85.32 	&	0.24 	\\
\midrule	
NASLung &	\textbf{90.77}	&	\underline{85.37} 	&	\underline{95.04} 	&	\textbf{89.29} 	&	16.84 	\\
\bottomrule												
\end{tabular}
\end{table}

\section{Conclusion}
\label{sec:conclusion}

In this paper, we develop an deep pulmonary nodule classifier with excellent accuracy/speed trade-off, by incrementally using 3D NAS, CBAM, A-Softmax loss, and the ensemble strategy. All the proposed techniques are demonstrated effective. Besides, the reasoning process is partially explainable. 
The learned attention maps provide a fine-grained modelling of the relationships between visual content (i.e. patterns of nodules) and clinical symptoms (e.g. lobular border and heterogeneous growth rate). 
Based on such success, it is promising to further explore the relationships between multiple nodules, as well as the changes of nodules during a period, to further boost the diagnosis accuracy and justify the diagnosis process \cite{yang2020relational}. 
Besides, it is critic to further explore the relational learning among mutli-modal data \cite{zhan2019exploring}, e.g. phenomics, dialogues, and reports, for improving the precision and interpretability of CAD systems \cite{xu2019end}. 
Finally, most NAS algorithms need a large number of labeled samples and a long training period. In medical imaging tasks, there are typically a small number of labeled samples. It is thus significant to develop few-shot NAS algorithms, which can learn optimal neural architectures based on few number of samples.  

\section*{Acknowledgements}
This work was supported in part by the National Natural Science Foundation of China under Grants 61971172, 61836002, 61702145, 61971339, and 61702143, in part by the Zhejiang Provincial Science Foundation under Grants Y18H160029, in part by the National Key R\&D Program of China under Grant 2018AAA0100603, and in part by the China Post-Doctoral Science Foundation under Grant 2019M653563.


\bibliography{mybibfile}

\end{document}